\documentclass[floats,twocolumn,prd,preprintnumbers,shortbibliography,nofootinbib,10pt]{revtex4-1}
\usepackage[colorlinks=true,urlcolor=blue,linkcolor=blue,citecolor=blue]{hyperref}
\usepackage{slashed,color,amsmath,amssymb,mathrsfs}
\usepackage{amsfonts,epsfig,amssymb}
\usepackage{graphicx}
\usepackage{hyperref}
\usepackage{longtable}
\usepackage{array}
\usepackage{accents}
\usepackage{tensor}
\usepackage{footmisc}
\usepackage{longtable}
\usepackage{booktabs}
\usepackage{multirow}
\usepackage{bm}
\usepackage{subfigure}
\usepackage{chngpage}

\newcommand{\sectitle}[1]{\vspace{.6cm}{\em #1.--}}
\allowdisplaybreaks

\begin{document}
\title[]{Nature of $X(2370)$}
\author{Xiang Sun$^{1,2}$}
\author{Ling-Yun Dai$^{1,2}$}\email{dailingyun@hnu.edu.cn}
\author{Shi-Qing Kuang$^{1,2}$}
\author{Wen Qin$^{1,2}$}\email{wqin@hnu.edu.cn}
\author{A. P. Szczepaniak$^{3,4,5}$}\email{aszczepa@indiana.edu}

\affiliation{$^{1}$\small School of Physics and Electronics, Hunan University, Changsha 410082, China\\
$^2$ Hunan Provincial Key Laboratory of High-Energy Scale Physics and Applications, Hunan University, Changsha 410082, China\\
$^3$ Physics Department, Indiana University, Bloomington, IN 47405, USA\\
$^4$ Center for Exploration of Energy and Matter, Indiana University, Bloomington, IN 47403, USA\\
$^5$ Thomas Jefferson National Accelerator Facility, Newport News, VA 23606, USA
}

\begin{abstract}
We address the nature of the $X(2370)$ resonance  observed in the $J/\psi$ radiative decays,  $J/\psi\rightarrow\gamma K^{+} K^{-}\eta'$, $J/\psi\rightarrow\gamma K_S K_S\eta'$ and  $J/\psi\rightarrow\gamma \pi^{+}\pi^{-}\eta'$. By studying the invariant mass spectra we confirm that decays of the $X(2370)$ into three pseudo-scalars  are well described by an effective chiral Lagrangian. We extract the branching ratio of $J/\psi\to X(2370)\gamma$ and show that it is an order of magnitude larger compared to the glueball production rate predicted by lattice QCD. This indicates  that $X(2370)$  is not likely to be a glueball candidate.
\end{abstract}


\maketitle

\sectitle{Introduction}
\label{Sec:I}
 Over the past decade important discoveries have been made  about strong interactions especially in what regards  the
  spectrum of hadrons.  The observation of charged mesons and baryons with hidden charm, {\it e.g.,}  the  $Z_c$ \cite{Ablikim:2013mio,Liu:2013dau}, and $P_c$ states, \cite{Aaij:2015tga,Aaij:2019vzc}
  indicates the possibility of  existence of compact multi-quark bound states that cannot be explained by the quark model.
Furthermore, there is a growing evidence that gluons, besides confining quarks,  can also act as hadron constituents resulting in
quark-gluon {\it aka}  hybrid states or pure glue made glueballs. For a recent review of the hybrid meson signatures and the
phenomenological studies addressing the role of gluons as constituents of hadrons
see, for example  \cite{Ketzer:2019wmd,Shepherd:2016dni,Woss:2020ayi,Ochs:2013gi,Bass:2018xmz,Szczepaniak:2001rg,JPAC:2018zyd,Bibrzycki:2021rwh} and references therein.

Recent analysis of the BESIII data on $J/\psi$ radiative decays to two pseudo-scalars have identified a multitude of  iso-scalar states and argument have been put forward that there is a
colorless, C-even pure glueball  among them \cite{Sarantsev:2021ein,Klempt:2021nuf,Rodas:2021tyb} Furthermore, the recent observation of the odderon \cite{Abazov:2020rus}, in the high-energy $pp$ and $p\bar p$ collisions may be related to existence of a  C-odd glueball resonance in the direct channel. It is the gluon compound in $J/\psi$ radiative decaying to three pseudo-scalars that we address in this paper.
Lattice QCD (LQCD)  predictions of  \cite{Morningstar:1999rf,Chen:2005mg, Gregory:2012hu,Sun:2017ipk,Gui:2019dtm} place a pseudo-scalar glueball
 mass above 2 GeV.  For example the most recent computation from   \cite{Gui:2019dtm} gives  $M_G\simeq2395\pm14$~MeV. However, since these are quenched calculations it is difficult to access the systematic uncertainties.

Glueballs are expected to be produced in radiative decays of the $J/\psi$~\cite{Close:2005vf} because  annihilation of the  $c\bar c$ pair 
 leaves behind a gluon rich component of the $J/\psi$ wave function. 
  Recently, the BESIII collaboration reported several, high statistics measurements of exclusive $J/\psi$ radiative decays.
 A structure with mass around 2.37~GeV, referred to as the $X(2370)$ has been seen in,  $\pi^{+}\pi^{-}\eta'$~\cite{Ablikim:2010au}
 and $\bar{K}K\eta'$ \cite{Ablikim:2019zyw} invariant mass  distributions.
In the former the mass and width of the $X(2370)$ are measured to be  $M = 2376.3 \pm 8.7(stat)^{+3.2}_{-4.3}(syst)$~MeV
 and $\Gamma= 83 \pm 17(stat)^{+44}_{-6}(syst)$~MeV,  with the statistic significance of 6.4$\sigma$,
 while in the latter decay, it was found that  $M = 2341.6 \pm 6.5(stat)\pm 5.7(syst)$~MeV and $\Gamma= 117 \pm 10(stat)\pm 8(syst)$~MeV
with the   8.3$\sigma$ statistical  significance. 
Notice that the $\eta(2225)$ has a comparable mass and width, but it has been seen in the $J/\psi \rightarrow \gamma \phi \phi$ \cite{BES:2008ecl}, decaying into four kaons. It is not our intention to examine all glueball candidates. Instead we focus  here on the study of the three-pseudoscalar meson spectrum given the recent measurements of several such 
channels by BESIII  and presence of a clear resonance signal 
in this spectrum.

There is no first principle method that would enable to distinguish a glueball from other inner components of a physical resonance. The two, however, are expected to have different phenomenological consequences and here  we propose  to compare the measured $J/\psi$ radiative decay branching ratios to those  predicted by
LQCD to investigate if the recently observed $X(2370)$ resonance is a good candidate for the pseudo-scalar glueball.
The $X(2370)$ has been considered in previous studies
\cite{Eshraim:2012jv,Qin:2017qes} where it was concluded that it may indeed correspond to the
pseudo-scalar  glueball  that was found in the LQCD simulations.  To verify this interpretation, it is necessary, however,  to consider the  production characteristics.
Following \cite{Eshraim:2012jv} we postulate
 the effective interactions between the $X(2370)$ and the  light scalar and pseudo-scalar  meson resonances
 that appear to dominate its decay spectrum.  The subsequent decays of these resonances
  are well studied, {\it e.g.} using the chiral theory   \cite{Dai:2019lmj}.
Combining these processes we construct a reaction
 model  to describe the $J/\psi$ radiative decays
 for all the  measured channels that contain the $X$,  $J/\psi\rightarrow\gamma K^{+} K^{-}\eta'$, $J/\psi\rightarrow\gamma K_S K_S\eta'$, $J/\psi\rightarrow\gamma \pi^{+}\pi^{-}\eta'$ and $J/\psi\rightarrow\gamma \eta\eta\eta'$ .
 With the branching ratios, ${\rm Br}[X\to PPP]$ of $X$ decaying into three pseudo-scalars  fixed by the reaction model and the branching ratio for the $J/\psi$ radiative decay in the mass region of the $X$,  ${\rm Br}[J/\psi\to\gamma X \to\gamma \bar{K} K\eta']$ determined by the experiment \cite{Ablikim:2019zyw},
 we  extract the branching ratio ${\rm Br}[J/\psi\to\gamma X]$
\begin{equation}
{\rm Br}[J/\psi\to\gamma X ] = \frac{
{\rm Br}[J/\psi\to\gamma X \to \gamma PPP]}{{\rm Br}[X \to PPP]} \label{eq:Br}
\end{equation}
By comparing our results with those of the  QCD predictions \cite{Gui:2019dtm} and the models  \cite{Eshraim:2012jv,Qin:2017qes},
we can therefore distinguish whether the $X(2370)$ is more likely to be a  glueball or a $q\bar q$ resonance.

\sectitle{Formalism}\label{Sec:II}
The effective Lagrangian for  $J/\psi\rightarrow\gamma X$, constrained by chiral and discreet symmetries, {\it e.g.,} charge conjugation and parity, can be written as
\begin{equation}
\mathcal{L}=g_{\gamma} \epsilon^{\mu\nu\alpha\beta} D_{\mu}\psi_{\nu} F_{\alpha \beta} X\,,  \label{equ:2}
\end{equation}
where $F_{\alpha\beta}$ is the usual electromagnetic field strength tensor.
Following \cite{Eshraim:2012jv,Eshraim:2019sgr,Eshraim:2020zrs} we apply the chiral effective  theory to describe the interactions  between the $X$ and nonets of scalar,
 $S$  and pseudo-scalar, $P$ fields  which is  given by,
\begin{align}
\mathcal{L}=ig_{X}X(det \Phi -det\Phi^{\dagger}) \,. \label{eq:L;XPhi}
\end{align}
 with $\Phi=\Phi_0+Z_S S(x) +i Z_P P(x)$, where $\Phi_0$ is a constant matrix
 that contains various condensates and $Z_S$, $Z_P$ are the wave function renormalization constants, 
For the pseudo-scalar nonet, $P$, we take the lightest $\pi$, $K$ mesons, and include the $\eta$ and $\eta'$ in the standard way as a result of mixing between the  $\eta_0$ and $\eta_8$~\footnote{For $\eta-\eta'$ mixing, as discussed in \cite{Dai:2017tew}, it is not clear that the double angles mixing scheme \cite{Guo:2011pa,Chen:2012vw} will  improve the analysis of $\eta'\to\pi^+\pi^-\gamma$. Thus we simply use one angle mixing scheme as what is done in Ref.\cite{Dai:2013joa,Qin:2020udp}. For mixing between $\eta,\eta'$ and gluon component, it is ignored here but can be found in Refs.~\cite{Khlebtsov:2020rte,Wang:2015kra,Mathieu:2009sg,Gutsche:2009jh}. }. 
Decays of $X\to KK\eta$ to heavier $\eta$ mesons have smaller phase space and their mixing with the lighter $\eta$'s is small,  thus they will be ignored in our analysis.  
 For the scalar nonet, $S$, there are two sets of  resonances with mass below 2~GeV \cite{Jaffe:2004ph} that are relevant. 
 The lighter set is associated with $ \{ \sigma, \kappa, a_0(980), f_0(980) \}$, and the heavier one with the  $ \{f_0(1370),K_0^*(1430),a_0(1450),f_0(1500) \}$  resonances.   There is  phenomenological evidence  that the states in the heavier set are dominated by $\bar{q}q$ configurations  \cite{Dai:2019lmj}, though a mixing with a glueball can not be neglected. The structure of the lighter scalars is still a mystery, but they have non-ignorable valance quark  components~\cite{Dai:2011bs,Dai:2014zta,Pelaez:2015qba,Yao:2020bxx}.  Out of these we construct two scalar nonets  $S=S_L,S_H$. Specifically,  we use the heavier iso-scalar, scalar mesons $f_0(1370)$,  $f_0(1500)$ and the  $f_0(1710)$,  mixed according to the model of \cite{Dai:2019lmj} to extract the two iso-scalar elements of the heaver multiplet $S= S_H$ as well as the scalar glueball $G$\footnote{ Notice that the scalar glueballs, mixing with $f_0$ states, are also discussed in Refs.~\cite{Fariborz:2021gtc,Fariborz:2018tyi,Fariborz:2018xxq,Leutgeb:2019lqu,Souza:2019ylx}:
 }
 \begin{equation}
    \label{eq:diag}
    \left( \begin{array}{c} \sigma_{n} \\ \sigma_{s} \\ G \end{array} \right) \; = \;\left( \begin{array}{ccc}
       \sqrt{1/3}   &   \sqrt{2/3}     &0   \\
       -\sqrt{2/3}    &  \sqrt{1/3}    &0  \\
           0              &  0         &1
                      \end{array} \right)  A \; \left( \begin{array}{c} f_0(1370)  \\ f_0(1500)  \\
               f_0(1710)  \end{array} \right)
    \,. \nonumber
    \end{equation}

where  the $3\times 3 $ mixing  matrix $A$ is given by  in 
  \cite{Dai:2019lmj}.
The other components of $S_L$ and $S_H$ originate from mixing between the physical states from the lighter set of resonances, $a_0(980)$ and $\kappa$ and the  $a_0(1450)$ and the $K_0^*(1430)$ for the heavier ones. They are given as \cite{Black:1999yz,Dai:2019lmj}
  \begin{eqnarray}
    a_{0,H}&=&  \cos \varphi_a a_0(1450)- \sin \varphi_a a_0(980) \, , \nonumber \\
    K_{0,H}^*&=&  \cos \varphi_k K_0^*(1430)- \sin \varphi_k K_0^*(700) \, . \nonumber
    \end{eqnarray}
The decays of the two nonets of the scalar resonances is described by the chiral effective Lagrangian from \cite{Dai:2019lmj},
\begin{align}
				\nonumber
		\mathcal{L}\;=&c_{d}^{H} \langle S_{H} u_{\mu} u^{\mu} \rangle + c_{m}^{H} \langle S_{H} \chi_{+} \rangle + \alpha_{H} \langle S_{H} u_{\mu} \rangle \langle u^{\mu} \rangle  \\
				\nonumber
				   &+\beta_{H}\langle S_{H}  \rangle \langle u_{\mu} u^{\mu} \rangle +\gamma_{H} \langle S_{H}  \rangle \langle u_{\mu}  \rangle \langle u^{\mu}  \rangle     \\
   &+c'_{d} G \langle u_{\mu} u^{\mu} \rangle +c'_{m} G \langle \chi_{+}\rangle +\gamma' G \langle u_{\mu} \rangle \langle  u^{\mu} \rangle\,,
				\nonumber \\				
  	&+ c_{d}^{L} \langle S_{L} u_{\mu} u^{\mu} \rangle + \alpha_{L} \langle S_{L} u_{\mu} \rangle \langle u^{\mu} \rangle +c_{m}^{L} \langle S_{L} \chi_{+}\rangle  \,. \label{eq:L;S}
\end{align}
The interactions among the pseudo-scalars are described by the chiral Lagrangian taken from \cite{Gasser:1983yg,Ecker:1989yg},  $\mathcal{L}= F^2 \langle u_{\mu}u^{\mu}+\chi_{+} \rangle/4$.
Notice that the vector meson resonances, such as $\rho(770)$, $\omega(782)$, and $\phi(1020)$, do not appear as  in the decay of the $X(2370)$.  The reason is that the $XVP$ vertex violates the C parity conservation.  Similarly, the axial vectors such as $a_1(1260)$, $b_1(1235)$ do not appear as the intermediate states because $APP$ is not allowed by the parity conservation.

As mentioned in the introduction, we need to extract the  radiative decay width of $J/\psi\to\gamma X$ to judge whether the $X(2370)$ is dominated by the glueball.
The experiment \cite{Ablikim:2019zyw} measured the product of the branching ratios
 ${\rm Br}[J/\psi\to\gamma X] {\rm Br}[X \to PPP]$.
The total width of the $X(2370)$, irrespective of its nature, can be estimated by summing over all the decay channels with three light pseudo-scalar final states,
 \begin{align}
	\Gamma_{X}(Q^2)= \sum_i 	\Gamma_{X \to (PPP)_i}, \label{eq:allwidth}
	\end{align}
with the sum ruining over
$KK\pi$, $\pi\pi\eta$, $KK\eta$,  $KK\eta'$,  $\pi\pi\eta'$  $\eta\eta\eta$,   $\eta\eta\eta'$, $\eta\eta'\eta'$  \cite{Eshraim:2012jv}.
Since  ${\rm Br}[X \to PPP]$ is given by a ratio of partial to total widths, the dependence on $g_X$ cancels and it can be predicted directly from the amplitudes constructed from the effective Lagrangians  in Eqs.~(\ref{eq:L;XPhi},\ref{eq:L;S}).
 Thus the  branching ratio   ${\rm Br}[J/\psi \to \gamma X]$  for production of the $X(2370)$ can be extracted directly using Eq.~(\ref{eq:Br}) with the denominator  fixed by the dynamics governing the $X (\to S P)\to PPP$ decays and the numerator given by the experiment.

\sectitle{Results and discussions}
\label{Sec:III}
The experiment gave the branching ratios of the $J/\psi\to\gamma X\to \gamma K^{+}K^{-}\eta'$, $J/\psi\to\gamma X\to \gamma K_S K_S\eta'$ and upper limit of the branching
ratio of $J/\psi\to\gamma X\to\gamma \eta \eta \eta'$. To estimate ${\rm Br}[X \to PPP]$, one needs to fix the couplings in $S\to PP$.
The mixing angles of scalars and that of $\eta-\eta'$, as well as the coupling constants
in  Eqs.~(\ref{eq:L;S}) were determined in \cite{Dai:2019lmj} through analysis of $PP$ production. 
In that analysis the scalar resonances do not appear as isolated Breit-Wigner amplitudes but decay into $PP$ where the final state interactions (FSI) are taken into account. 
We perform two analyses in the present paper according to different ways to deal with $S\to PP$. In what we refer to as Sol.~I,
the parameters of  $S\to PP$ are taken from \cite{Dai:2019lmj}, and then they are input into the analysis of $J/\psi$ radiative decays. In contrast, in Sol.~II, the $S\to PP$ decays are refitted without the FSI. Note that the FSI of $\pi\pi$-$K\bar{K}$ has been partly restored by the Breit-Wigner forms of the scalars appear in the intermediate states. 
The results of ${\rm Br}[X\to PPP]$ are  shown in Table \ref{tab:X;Br}.
\begin{table}[htbp]
\begin{ruledtabular}
\begin{tabular}{lcclcc}
                       & Sol.~I                           & Sol.~II                                                  \\ \hline
  ${\rm Br}[X\to KK\eta']$          & $0.981\pm0.05\times10^{-2}$    & $0.983\pm0.04\times10^{-2}$     \\
  ${\rm Br}[X\to KK\pi]$       & $0.522\pm0.01$    & $0.525\pm0.01$          \\
  ${\rm Br}[X\to \pi\pi\eta']$     & $2.64\pm0.13\times10^{-2}$   & $2.71\pm0.16\times10^{-2}$   \\
  ${\rm Br}[X\to KK\eta]$      & $6.73\pm0.13\times10^{-2}$   & $6.55\pm0.13\times10^{-2}$      \\
  ${\rm Br}[X\to \eta\eta\eta']$   & $3.27\pm0.26\times10^{-3}$  &
  $4.62\pm0.23\times10^{-4}$   \\
  ${\rm Br}[X\to \pi\pi\eta]$        & $0.356\pm0.01$      & $0.359\pm0.01$           \\
  ${\rm Br}[X\to \eta\eta'\eta']$  & $0.0$       & $0.0$  \\
  ${\rm Br}[X\to \eta\eta\eta]$    & $1.56\pm0.05\times10^{-2}$   & $1.28\pm0.04\times10^{-2}$     \\
  
\end{tabular}
\end{ruledtabular}
\caption{Predictions of branching ratios of the $X(2370)$ decays. }\label{tab:X;Br}
\end{table}
The results of Sol.~I are only a slightly  different from that of Sol.~II, except for the ${\rm Br}[X\to \eta\eta\eta']$.  The latter
 is caused by the difference in  $f_0(1370)\to \eta\eta$ decay widths found between Sol.~I and Sol.~II with the former
  almost 4 times larger. 
From these branching ratios,  it is easy to compute  the branching ratios of the $J/\psi\to\gamma X$ directly from the experimental results for  ${\rm Br}[J/\psi\to\gamma X\to \gamma PPP]$ through Eq.~(\ref{eq:Br}).
In practice, to do so  we perform a combined fit
 of ${\rm Br}[J/\psi\to\gamma X \to \gamma K^{+} K^{-}\eta']$, ${\rm Br}[J/\psi\to\gamma X \to\gamma K_S K_S\eta']$ and impose the upper limit of ${\rm Br}[J/\psi\to\gamma X \to\gamma \pi^{+}\pi^{-}\eta']$ to extract the single parameter,  ${\rm Br}[J/\psi \to \gamma X]$,   The results are also named as Sols.~I and II according to different $S\to PP$ inputs and are shown in Table.\ref{tab:BR}.  
\begin{table}[htbp]
{\footnotesize
\begin{tabular}{lccc}
\hline\hline
                                    & Sol. I              & Sol. II             &  Results from exp. or LQCD  \\ \hline
  ${\rm Br}^{(1)}(10^{-5})$   & $1.45\pm0.23$    & $1.99\pm0.38$       &   $1.79\pm0.23\pm0.65$  \cite{Ablikim:2019zyw}  \\
  ${\rm Br}^{(2)}(10^{-5})$ & $0.68\pm0.11$    & $0.94\pm0.18$         &   $1.18\pm0.32\pm0.39$  \cite{Ablikim:2019zyw}  \\
  ${\rm Br}^{(3)}(10^{-6})$     & $9.20\pm1.26$   & $1.79\pm0.25$        &    $<9.2$ \cite{Ablikim:2020bnb}  \\
  ${\rm Br}^{{\rm tot}}(10^{-3})$   & $2.87\pm0.68$    & $3.95\pm0.71$      & $0.231\pm0.090$ \cite{Gui:2019dtm}          \\
  \hline\hline
\end{tabular}
\caption{Predictions of branching ratios of $J/\psi$ radiative decays from our fit. Here the superscripts \lq 1,2,3' represent for the processes of $J/\psi\to\gamma X\to\gamma K^{+}K^{-}\eta'$, $J/\psi\to\gamma X\to\gamma K_S K_S\eta'$ and $J/\psi\to\gamma X\to\gamma \eta\eta\eta'$, respectively. The label \lq tot' is for the process of $J/\psi\to\gamma X(2370)$. The $\chi^2_{d.o.f}$ is 0.15 and 0.60 for Sol.~I and Sol.~II, respectively. }\label{tab:BR}
}
  \end{table}
As can be found, when the fitted value for  ${\rm Br}[J/\psi\to\gamma X]$ is used to  compute  ${\rm Br}[J/\psi\to\gamma X\to \gamma PPP]$, the result agrees with the experiment within the experimental uncertainties.
  In Sol.~I, and II, we find
      ${\rm Br}[J/\psi\to\gamma X]= 2.87\pm0.68 \times 10^{-3}$ and
  ${\rm Br}[J/\psi\to\gamma X]=3.95\pm0.71\times 10^{-3}$, respectively,
while in quenched LQCD, the pure glueball production rate is found to be
${\rm Br}[J/\psi\to\gamma X]=0.231\pm0.090\times 10^{-3}$ \cite{Gui:2019dtm}, for $M_{G}=2.395$~GeV.
Our result  is almost one order larger than that of LQCD
and it implies that the glueball can not be the dominant component of the $X(2370)$.

To further study the nature of the $X(2370)$ and as a check of the amplitude model, we also perform a combined analysis of the branching ratios and the invariant mass spectra in $J/\psi\to \gamma K^{+} K^{-}\eta'$, $J/\psi\to\gamma K_S K_S\eta'$ and  $J/\psi\to\gamma \pi^{+}\pi^{-}\eta'$. 
The differential decay rate is given as
\begin{eqnarray}
  \frac{d\Gamma_{J/\psi \to \gamma (PPP)_i}}{dQ}&=&\frac{(M_{J/\psi}^2-Q^2)}{128 Q M^3_{J/\psi}(2\pi)^5} \nonumber\\
 && \int ds \int dt~ \vert \mathcal{M}^{(i)}_{X}+ \mathcal{M}^{(i)}_{b.g.} \vert^2\,. \label{eq:dGdQ;0}	
\end{eqnarray}
Here $\mathcal{M}^{(i)}_{X}$ comes from Chiral effective field theory, describing the amplitudes of $J/\Psi\to\gamma X\to\gamma (PPP)_i$, where \lq $i$' denotes the $i$-th channel.  $\mathcal{M}^{(i)}_{b.g.}$ is a background (not from the intermediate state  X(2370) ) parametrized by first-order polynomials of $s$. 
This distribution is fitted to the (unnormalized)  invariant  mass spectrum and one can obtain the branching ratio of $J/\Psi \to \gamma X \to \gamma (PPP)_i$  by integrating Eq.(\ref{eq:dGdQ;0}) under the peak ({\it i.e.} with background removed). 

\begin{figure}[htbp]
\centering
\subfigure[]{\label{fit21}\includegraphics[width=0.23\textwidth,height=0.13\textheight]{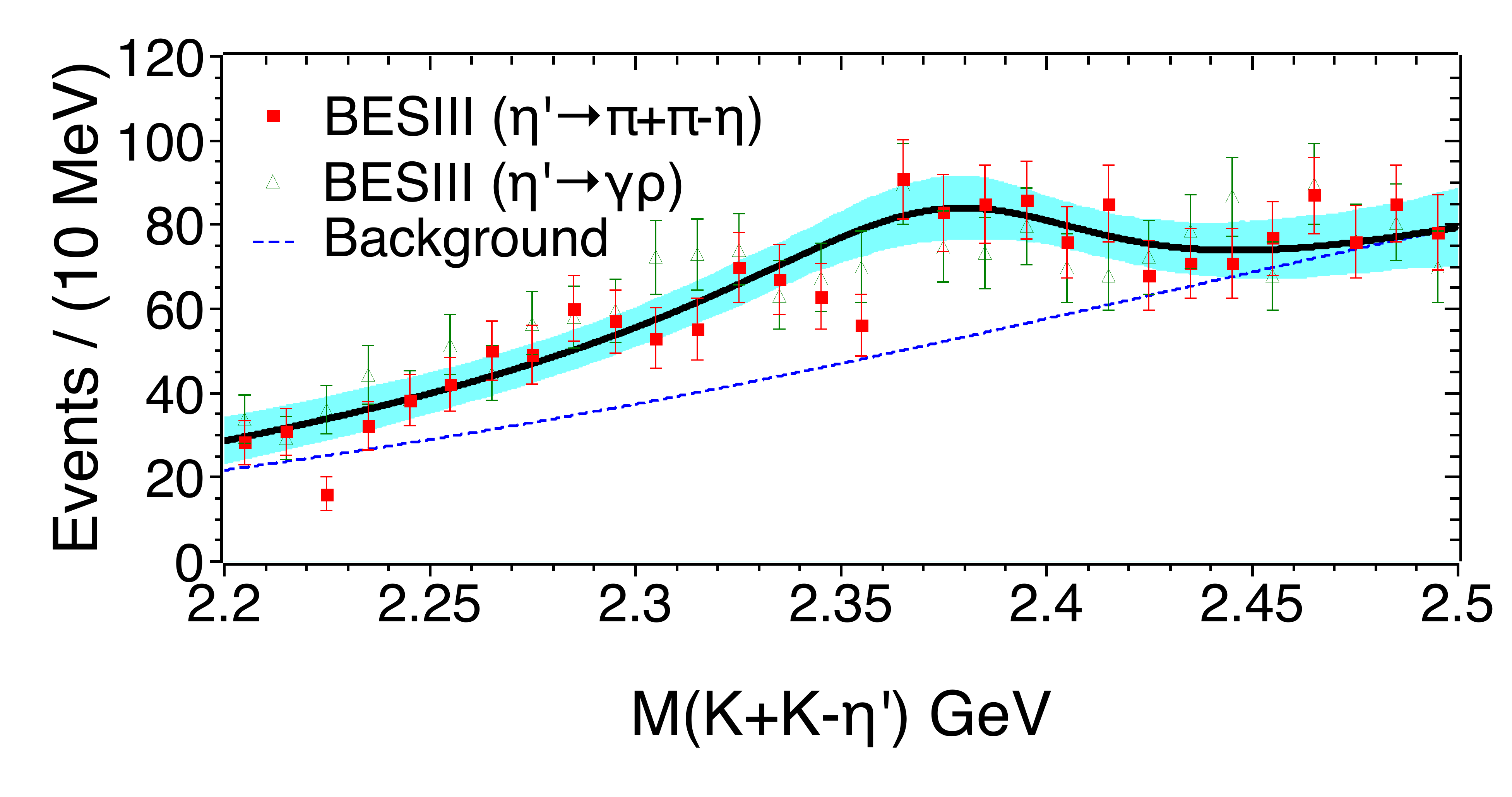}}
\subfigure[]{\label{fit23}\includegraphics[width=0.23\textwidth,height=0.13\textheight]{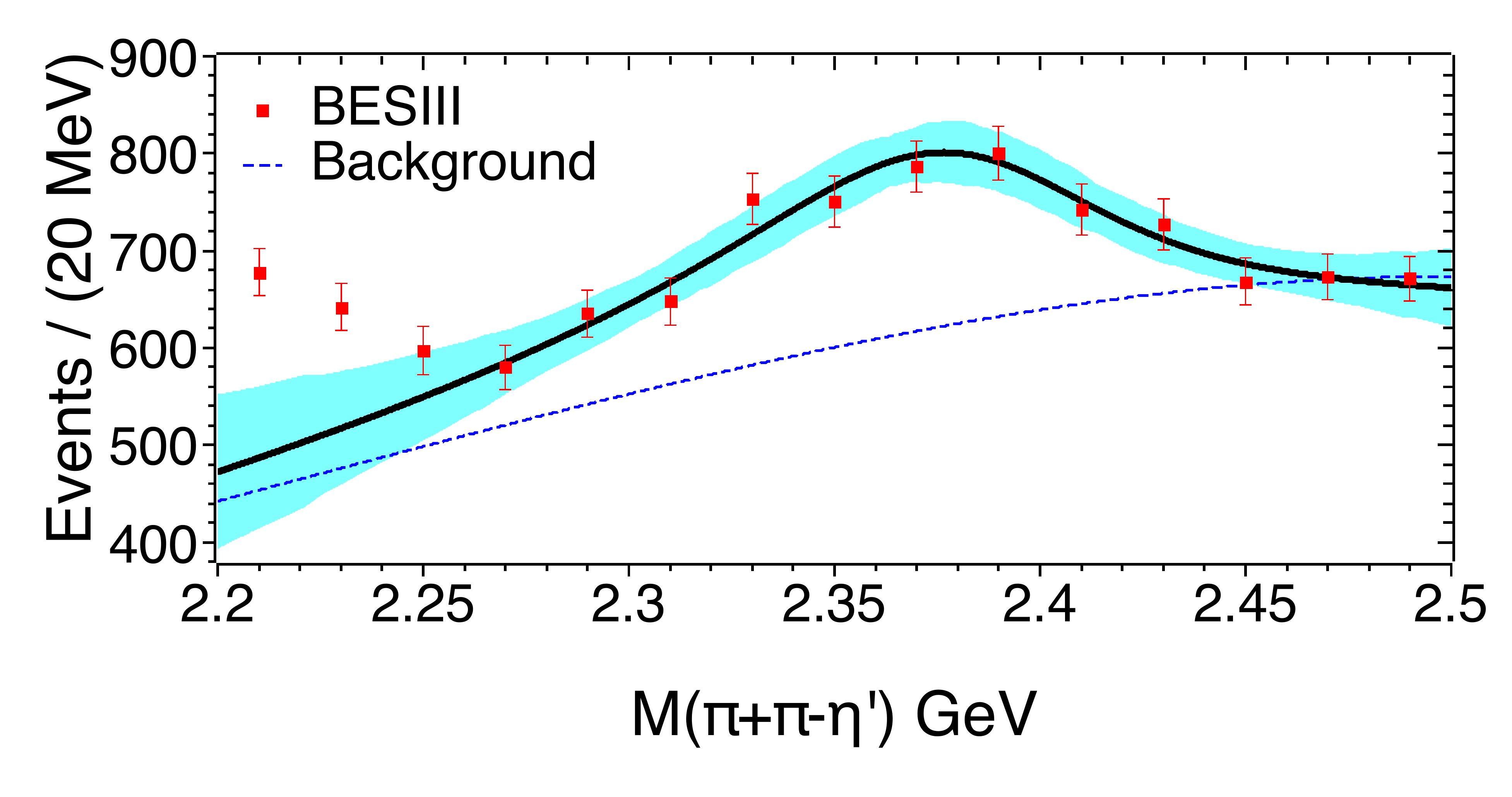}}
\caption{Combined fit  results of
$ K^{+} K^{-}\eta'$ and  $ \pi^{+} \pi^{-}\eta'$ invariant mass distributions.
The experimental data displayed are from BES\uppercase\expandafter{\romannumeral3} collaboration,
\cite{Ablikim:2019zyw} for the $J/\psi\rightarrow\gamma K^{+} K^{-}\eta'$, and \cite{Ablikim:2010au} for the $J/\psi\rightarrow\gamma \pi^{+}\pi^{-}\eta'$, respectively.
Two $\eta'$ decay modes, {\it i.e.} $\eta'\to\pi^+\pi^-\eta$ and $\eta'\to \gamma \rho^0$ are displayed for
$J/\psi\rightarrow\gamma K^{+} K^{-}\eta'$  process with red filled square and green open triangle.
Our resluts are shown by the black solid lines with cyan bands. The backgrounds are shown with blue dotted lines. }
\label{Fig:events}
\end{figure}

In particular in Fig.\ref{Fig:events} we show a sample fit result obtained for $J/\psi\rightarrow\gamma K^{+} K^{-}\eta'$ and $J/\psi\rightarrow\gamma \pi^{+} \pi^{-}\eta'$ mass spectra. 
In our analysis we also include 
 backgrounds, which are caused by processes where thresholds for production of intermediate states are far away from the $X(2370)$ region, for instance, $J/\psi\to\gamma \eta (\eta')\to\gamma \bar{K}K\eta'$. These contribute with a smooth function of the invariant mass and when subtracted
  form the overall intensity the contribution of the $X(2370)$ is found to be very well described
  by a  Breit-Wigner resonance with  mass and width   $2387.8\pm1.3$~MeV, and $119.6\pm6.7$~MeV, respectively.
These values are close to those of the experimental analysis from  \cite{Ablikim:2019zyw,Ablikim:2010au}.
Finally, the combined analysis gives ${\rm Br}[J/\psi\to\gamma X]=3.26\pm0.81\times 10^{-3}$, very close to that of Sol.~II. 

\sectitle{A phenomenological analysis of the glueball production rate}\label{sec:physical;G}
As is known the glueball is a pure state (G), but it  mixes with quark components to form a physical  states, $X$.  Based on $U(1)_A$ anomaly, the dominant underlying mechanism of pseudo-scalar production in  $J/\psi$ radiative decay is via $c \bar c$ annihilation into two gluons and a photon \cite{Chen:2014yta,Fu:2011yy,Zhao:2010mm}, and the production rate fraction of the physical state $X$ and pseudo-scalar meson $\eta$ can be expressed as \cite{Qin:2017qes},
  \begin{equation}\label{production-fraction}
  \frac{{\rm Br}[J/\psi\to\gamma X]}{{\rm Br}[J/\psi\to\gamma \eta]}=\left(\frac{\alpha_{X}}{\alpha_{\eta}}\right)^2\left(\frac{M^2_{J/\psi}-M^2_X}{M^2_{J/\psi}-M^2_\eta}\right)^3.
  \end{equation}
 Here  $\alpha_i$, ($i=X,\eta$) stand for the matrix elements
   $\langle 0|  \alpha_s G_{\mu\nu} \tilde{G}^{\mu\nu}|i \rangle$, $G_{\mu\nu}$  and $\tilde{G}^{\mu\nu}$ denote the gluon field strength tensor and its dual, respectively. 
We first assume that $X$ is dominated the glueball, {{\it i.e.} $|X \rangle = |G\rangle$}. Then we can use the pure gauge lattice results from  \cite{Gui:2019dtm}  for ${\rm Br}[J/\psi\to\gamma G]=0.231\pm0.090\times 10^{-3}$ and for $\alpha_X=\alpha_g=-0.054 \mbox{ GeV}^3$\cite{Zyla:2020zbs}. Then using  the experimental data for  ${\rm Br}[J/\psi \to \gamma \eta] = 1.11\pm0.03\times 10^{-4}$ \cite{Zyla:2020zbs}, from  Eq.~(\ref{production-fraction}) we obtain  $\alpha_\eta=0.031 \mbox{ GeV}^3$. 
Now that we have determined $\alpha_\eta$, we input this value into the model of \cite{Qin:2017qes}, which enables to predict the physical matrix elements $\alpha_X$ for the physical state,  
that is a mixture of the glueball and the 
 $q\bar q$ component.  If this state was  to be identified with the $X$, we can determine using  Eq.~(\ref{production-fraction}) the production rate,  obtaining ${\rm Br}[J/\psi \to \gamma X] =0.487\pm 0.143\times 10^{-3}$.
When compared  with the result from our analysis of the experimental data, this further confirms our conclusion that the $X(2370)$ is not likely to be a glueball.

\sectitle{Summary}
\label{Sec:IV}
In this work we constructed  $J/\psi\rightarrow\gamma K^{+} K^{-}\eta'$, $J/\psi\rightarrow\gamma K_S K_S\eta'$, $\gamma\pi^{+}\pi^{-}\eta'$, and $\gamma\eta\eta\eta'$ decay amplitudes using chiral effective Lagrangian.
 It is found that the ${\rm Br}[J/\psi \to \gamma X]$ can be directly extracted from the experiment measurement of ${\rm Br}[J/\psi\to\gamma X \to \gamma PPP]$, where the ${\rm Br}[X \to PPP]$ is fixed by these amplitudes.
The branching ratio of ${\rm Br}[J/\psi \to X \gamma  ]$ is
found to be $2.87\pm0.68\times 10^{-3}$ or $3.95\pm0.71\times 10^{-3}$ depending  on the choice of parameters
and  are one order of magnitude  larger than that of LQCD, $0.231\pm0.090\times 10^{-3}$, or $0.487\pm 0.143\times 10^{-3}$ from a LQCD motivated phenomenology analysis. This is confirmed by a refined analysis of the amplitude model that  also considers the invariant mass spectra.
Our result is a strong evidence that the $X(2370)$ is not dominated by a glueball component.
Future experiments by BESIII at BEPCII and BelleII at superKEKB, which focus on the branching ratios of $X$ decays, would be rather helpful to study the nature of the $X(2370)$.

\sectitle{Acknowledgements}
\label{Sec:V}
This work is supported by Joint Large Scale Scientific Facility Funds of the National Natural Science Foundation of China and Chinese Academy of Sciences (CAS) under Contract No.U1932110, National Natural Science Foundation of China with Grants No.11805059 and No.11675051,  Fundamental Research Funds for the central Universities and   U.S. Department of Energy Grants No. DE-AC05-06OR23177 and No. DE-FG02- 87ER40365.

\bibliography{ref}

\end{document}